\documentclass[prd,letterpaper,twocolumn,aps,showpacs,nofootinbib,nobibnotes,superscriptaddress,preprintnumbers,floatfix]{revtex4-1}

\usepackage{amsfonts,amssymb}

\usepackage[figuresright]{rotating}
\usepackage[colorlinks=true,linkcolor=red,urlcolor=blue,citecolor=blue]{hyperref}
\usepackage{bm,color,amsmath}
\usepackage{mathrsfs}
\usepackage{xcolor}
\usepackage{dcolumn}
\usepackage{multirow}
\usepackage{cancel}
\usepackage[normalem]{ulem}
\graphicspath{{./fig/}{./ps/}}
\usepackage{xfrac}

\def \pa{\partial}

\newcommand{\stonybrook}{Physics and Astronomy Department, Stony Brook University, Stony Brook, New York 11794, USA}

\begin{document}

\title{Mass-varying Dark Matter from a Phase Transition}

\author{Sayan~Mandal}
\affiliation{\stonybrook}

\author{Neelima~Sehgal}
\affiliation{\stonybrook}

\begin{abstract}
We propose a mass-varying dark matter (MVDM) model consisting of a scalar field and a fermionic field interacting via a simple Yukawa coupling, and containing an exponential self-interaction potential for the scalar field.
Analyzing the evolution of this coupled scalar-fermion system in an expanding Universe, we find that it initially behaves like radiation but then undergoes a phase transition after which it behaves like pressureless dark matter.  The one free parameter of this model is the temperature at which the phase transition occurs; the mass of the dark matter particle, given by the mass of the fermion, is derived from this. 
For a phase transition temperature between 10 MeV and $10^{7}$ GeV, the current dark matter relic density is achieved for a fermion mass in the range of 1 GeV to $10^{9}$ GeV. In this dark matter model, the scalar becomes a sub-dominant unclustered component of dark matter that can lower the amplitude of structure formation by up to a few percent. Another feature is that the mass-varying fermion component can lead to discrepant measurements of the current dark matter density of about ten percent inferred from early and late-time probes assuming $\Lambda$CDM.
\end{abstract}

\maketitle

\section{Introduction}
\label{secIntro}

Dark matter comprises a significant fraction of the matter-energy budget of the Universe~\cite{Planck:2018vyg}. Many models of dark matter have been proposed where the dark matter interacts with standard model particles only through the gravitational force, as well as interacts weakly with the standard model via mechanisms outside of gravity, e.g.~\cite{Boyarsky:2018tvu, Kopp:2021jlk, Arbey:2021gdg, Profumo:2019ujg, Kamionkowski:1997zb, Kamionkowski:1998is, Jungman:1995df, Fox:2013pia, Ferrer:2006hy, Bertone:2005xv, Davis:2015vla, Zacek:2007mi, Griest:1993cv, Freese:1985qw, Bertone:2004pz, Cheng:2002ej, Hooper:2004dc, Tucker-Smith:2001myb, Kaplan:1991ah, Farrar:2004qy, Turner:1983he, Feng:2008ya, Lee:2011jk, Zhou:2011fr, Belanger:2007dx, Enqvist:1988we, Borah:2011ve, Batell:2010bp, Chen:2009ab, Cui:2009xq, Arkani-Hamed:2008hhe, Arkani-Hamed:2008kxc, Kolb:1998ki, Kim:2006af, Cirelli:2005uq, Masip:2005fv, Pierce:2004mk, Mahbubani:2005pt, Arkani-Hamed:2005zuc, Arkani-Hamed:2006wnf, Carvajal:2021fxu, Berlin:2017ife, DAgnolo:2017dbv, DAgnolo:2018wcn, DAgnolo:2019zkf, Kramer:2020sbb, DAgnolo:2020mpt, Frumkin:2021zng, Pospelov:2007mp, DEramo:2010keq, Hochberg:2014dra, Smirnov:2020zwf, Kuflik:2015isi, Dodelson:1993je, Kusenko:2006rh, Petraki:2007gq, Shakya:2015xnx, McDonald:2001vt, Asaka:2005cn, Asaka:2006fs, Gopalakrishna:2006kr, Page:2007sh, Covi:2002vw, Cheung:2011mg, Kolda:2014ppa, Monteux:2015qqa, Benakli:2017whb, Co:2016fln, Hall:2012zp, Cheung:2011nn, Choi:2005vq, Moroi:1993mb, Co:2015pka, Bae:2014rfa, Essig:2015cda, Essig:2010ye, Essig:2011nj, Redondo:2008ec, Chu:2011be, Chu:2013jja, Chung:2001cb, Kuzmin:1998kk, Chung:1998is, Chung:1998ua, Ema:2018ucl, Fairbairn:2018bsw, Ema:2019yrd, Kolb:2020fwh, Graham:2015rva, Asadi:2022njl, Elor:2021swj, Baker:2016xzo, Baker:2017zwx, Baker:2018vos, Baker:2019ndr, Baker:2021zsf, Azatov:2021ifm, Azatov:2022tii, Daido:2017wwb, Daido:2017tbr, Croon:2022gwq, Croon:2020ntf, Chao:2020adk, Hong:2020est, Chway:2019kft, Cohen:2008nb, Salucci:2018hqu}. However, there has been no direct detection of dark matter in particle physics experiments to date, and cosmological observations suggest a good fit to $\Lambda$CDM with possible hints of deviations being inconclusive at this point.  Investigating the full theoretical landscape of viable dark matter models as broadly as possible may offer further clues to potential detection avenues.

In this work, we propose a model where a scalar field with a self-interaction potential is Yukawa-coupled to a fermion and gives the fermion an effective temperature-dependent mass via this interaction. The coupled fermion-scalar system considered here initially behaves like radiation, but undergoes a phase transition in the early Universe after which it behaves like pressureless dark matter. This model provides a natural way of explaining the mass of a heavy fermion dark matter particle. It also presents a novel mechanism of creating dark matter from a first-order phase transition as the Universe cools. While in this model dark matter only interacts with the standard model gravitationally, it provides two interesting cosmological features: 1)~the fermion mass increases slowly between the phase transition and today, and 2)~the scalar becomes a uniform rolling field after the phase transition, comprising a sub-dominant unclustered component of dark matter.  Our work repurposes the formalism of the mass-varying neutrino (MaVaN) model of~\cite{Fardon:2003eh}, later re-analyzed by~\cite{Chitov:2009ph, Mandal:2019kkv} in the context of finite-temperature field theory, which aimed to explain the non-zero masses of the standard model neutrinos from their interaction with quintessence dark energy; instead we adapt this framework to explain a heavy dark matter mass via an interaction with a scalar.

This paper is arranged as follows.  In Section~\ref{secFormalism}, we describe the model and its evolution through and after the phase transition. In Section~\ref{secPheno}, we discuss the phenomenological implications, and in Section~\ref{secDiscuss}, we elaborate on possible observational signatures, summarize, and conclude.  Throughout this work, we use the $(+,-,-,-)$ metric, work in natural units where $\hbar=c=k_B=1$, and set the present day relative abundance of dark matter $\Omega_{\rm DM,0} = 0.26$ to match the latest \textit{Planck} 2018 TT, TE, EE + lowE + lensing + BOSS BAO data~\cite{Planck:2018vyg}. Here and subsequently the subscript~$_0$ denotes the present value for time-dependent quantities.

\section{The model}
\label{secFormalism}

We consider a scalar field $\varphi$ with a self-interaction potential $U(\varphi)$ that governs its dynamics, and a fermion $\psi$.  We assume for simplicity a spatially flat Universe, but note that the dynamics of the model does not require flatness.  We describe the Universe by the Friedmann-Lema\^{i}tre-Robertson-Walker metric $ds^2=dt^2-a^2(t)d\mathbf{x}^2$, where $t$ is the cosmic time, $a(t)$ is the scale factor of the Universe, and $\mathbf{x}$ are the comoving coordinates.
The dynamics of the coupled scalar-fermion system is described by the action $S_{\varphi\psi}=\int d^4x\sqrt{-g}\mathcal{L}$, where $g\equiv\det(g_{\mu\nu})$, and $\mathcal{L}$ is the Lagrangian given by
\begin{equation}\label{eMainLag}
    \mathcal{L}=\frac{1}{2}\pa_\mu\varphi\,\pa^\mu\varphi-U(\varphi)+\bar{\psi}\left(i\cancel{\pa}-\bar{m}_\psi\right)\psi-g_Y\bar{\psi}\varphi\psi.
\end{equation}
Here, $\pa_\mu\varphi\,\pa^\mu\varphi/2$ is the kinetic energy of the scalar field, $i\bar{\psi}\cancel{\pa}\psi$ is the kinetic energy of the fermion using the notation $\cancel{\pa}\equiv\gamma^\mu\pa_\mu$ where $\gamma^\mu$ are the Dirac matrices, $\bar{\psi}\equiv\psi^\dagger\gamma^0$ is the antiparticle of $\psi$, $\bar{m}_\psi$ is the bare mass of the fermion, and $g_Y\bar{\psi}\varphi\psi$ is the Yukawa coupling between the fermionic and scalar fields.
We choose this coupling because it is the simplest Lorentz-invariant coupling between a fermion and a scalar.
In this work, we will set the bare mass $\bar{m}_\psi$ to be negligibly small, and ignore it from all subsequent expressions.  We also ignore higher-order terms involving $\psi$ that have dimension larger than four, since we will assume that they are irrelevant at the energy scales of interest for our model by suppressing them with small dimensionful coupling constants (and correspondingly heavy mass scales).  

Since the scalar field has no protective symmetries, it will couple with the Higgs field $H$ in the standard model via interactions of the form $g_H\varphi^2|H|^2$ and $G_H\varphi|H|^2$, which will make the fields of our model couple with standard model particles non-gravitationally. In Appendix~\ref{AddScalarInt}, we quantify the allowed strength of the couplings constants $g_H$ and $G_H$ in order to preserve the dark matter dynamics and find that these interactions can be made negligibly small. Thus, we ignore them from our analysis below.  However, we note that this required coupling ensures equilibrium between the standard model and the dark matter sector at some point in the early Universe, and thus an approximately common temperature between these two sectors.

The partition function describing the system is the path integral over all field configurations of the system, and is written conveniently by performing a change of variables to Euclidean time\footnote{This change of variables makes the metric signature Euclidean when written in terms of $\tau$, and the path integral is consequently easier to perform. It also makes a natural transition to studying the system at finite temperature (see also Footnote~\ref{FootFiniteT}).} $\tau=it$ as
\begin{equation}
    \mathcal{Z}_{\varphi\psi}=\int\mathcal{D}\varphi\,\mathcal{D}\bar{\psi}\,\mathcal{D}\psi\,e^{S^E_{\varphi\psi}},
\end{equation}
where $S^E_{\varphi\psi}$ is the action in terms of the variable $\tau$.
We can integrate out the fermion fields to write this in terms of an effective scalar action $S^E_{\rm eff}(\varphi)$,
\begin{equation}\label{eTotalPartFun-2}
\mathcal{Z}_{\varphi\psi}=\int\mathcal{D}\varphi\,e^{S^E_\mathrm{eff}(\varphi)}=\int\mathcal{D}\varphi\,e^{\left[S^E_\varphi+\ln\{\det\hat{D}(\varphi)\}\right]},
\end{equation}
where $S^E_\varphi$ is the scalar action written in terms of $\tau$ and
\begin{equation}\label{eDiracOp}
\hat{D}(\varphi)=-\beta\left[\frac{\pa}{\pa \tau}-i\frac{\gamma^0\bm{\gamma}\cdot\nabla}{a}+\gamma^0g_Y\varphi-\mu\right].
\end{equation}
Here $\mu$ is the chemical potential associated with the fermions, and $\beta=1/T$ is the inverse of temperature\footnote{\label{FootFiniteT} To analyse a system at finite temperature $T=1/\beta$, we evaluate the action by integrating from $\tau=0$ to $\beta$ with periodic (anti-periodic) boundary conditions for bosons (fermions) in $\tau\in[0,\beta]$.}.
For simplicity, we will set $\mu=0$, which means either the $\psi$ are Majorana fermions, or there are an equal number of Dirac fermions and anti-fermions in the $T=0$ ground state.
Letting the effective action $S^E_\mathrm{eff}(\varphi)$ be minimized when the scalar attains the value $\varphi=\varphi_m$, we can see from Eq.~\eqref{eDiracOp} that the fermion gets an effective mass of\footnote{After the phase transition, the effective action no longer has a minimum, and the solution $\varphi$ is obtained in a different way, as described in Sec.~\ref{SecAfterPT}.}
\begin{equation}\label{eFermMass}
m_\psi=g_Y\varphi_m.
\end{equation}
At this minimum point we can write $\mathcal{Z}_{\varphi\psi}=\mathcal{Z}_F\,e^{-\beta V U(\varphi_c)}$, where $\mathcal{Z}_F$ is the fermionic partition function\footnote{For the FLRW metric we have considered, $\sqrt{-g}=a^3$. Also, $V$ is the overall volume of the system which drops out of all subsequent expressions since we divide by the volume to obtain the free energy density and the pressure.},
\begin{equation}\label{eFermPartFunc}
    \mathcal{Z}_F=\int\mathcal{D}\bar{\psi}\,\mathcal{D}\psi\,\exp\left[\frac{1}{\beta}\int_0^\beta a^3 d\tau\int d^3\mathbf{x}\,\bar{\psi}\hat{D}_{\mu=0}(\varphi_m)\psi\right].
\end{equation}
This implies the free energy density $F_{\varphi\psi}\equiv-\ln\mathcal{Z}_{\varphi\psi}/\beta V$ of the combined system can be written as
\begin{equation}\label{eTotalFreeEn}
F_{\varphi\psi}(\varphi_m)=U(\varphi_m)+F_F(\varphi_m),
\end{equation}
with $F_F\equiv-\ln\mathcal{Z}_F/\beta V$ being the free energy density of the fermions,
\begin{equation}\label{eFFDef}
F_F=-\frac{1}{3\pi^2}\int_0^\infty\frac{dp\,p^4}{\epsilon(p)}\left[n_F(\epsilon_+)+n_F(\epsilon_-)\right],
\end{equation}
expressed in terms of the Fermi distribution function $n_F(x)=(e^{\beta x}+1)^{-1}$, with $\epsilon(p)=\sqrt{p^2+m_\psi^2}$ and $\epsilon_\pm=\epsilon(p)\pm\mu$.
The free-energy density $F_{\varphi\psi}$ is the effective thermodynamic potential for the scalar field, where $F_F(\varphi)$ is adding finite-temperature corrections from the scalar field's interaction with the fermion.
We note that the one-loop contribution to the effective potential at a finite temperature contains this temperature-dependent part $F_F(\varphi)$, as well as a zero-temperature contribution that we assume is already included in the Lagrangian in Eq.~\eqref{eMainLag}; we discuss this in more detail in Appendix~\ref{Zero-temp}.

The mass of the scalar field $\varphi$ is defined as
\begin{equation}\label{eRenormMass}
m_\varphi^2\equiv \left.\frac{\pa^2 F_{\varphi\psi}(\varphi)}{\pa\varphi^2}\right|_{\varphi=\varphi_m}.
\end{equation}
The minimization of the free energy density $F_{\varphi\psi}$ at $\varphi_m$ implies
\begin{equation}\label{eSaddleConditions}
\left.\frac{\pa F_{\varphi\psi}(\varphi)}{\pa\varphi}\right|_{\varphi=\varphi_m}=0,\quad \left.\frac{\pa^2F_{\varphi\psi}(\varphi)}{\pa\varphi^2}\right|_{\varphi=\varphi_m}>0.
\end{equation}
Using the first of these and Eq.~\eqref{eTotalFreeEn}, we obtain 
\begin{equation}\label{eMassEq}
\left.\left(\frac{\partial U(\varphi)}{\partial\varphi}+\frac{\pa F_F}{\pa \varphi}\right)\right|_{\varphi=\varphi_m}=0.
\end{equation}
Solving Eq.~\eqref{eMassEq} yields $\varphi_m$, and thus the mass of the fermion given by Eq.~\eqref{eFermMass}, prior to and at the phase transition.

\subsection{Choice of the scalar potential $U(\varphi)$}
\label{secPot}

The choice of $U(\varphi)$ is critical to the overall dynamics of this system.  We choose the scalar field to have an exponential potential, which can arise generically from Kaluza-Klein, superstring, supergravity, and higher-order gravity theories~\cite{Ferreira:1997hj, Wetterich:1984wd, Halliwell:1985bx, Cremmer:1983bf, Ellis:1983sf, Nishino:1984gk, Witten:1985xb, Dine:1985rz, Barrow:1988xh}.
This potential is given by
\begin{equation}\label{eFJPot}
U(\varphi) = M^4 e^{-{\sfrac{\lambda\varphi}{M}}},
\end{equation}
where $M$ and $\lambda$ are free parameters to be determined from the phase transition temperature and the relic abundance of dark matter.  As we will see below, the coupling constant $g_Y$ is not an independent parameter of our model because it always appears in combination with $\lambda$ as  $g_Y/\lambda$; thus in effect $g_Y/\lambda$ and $M$ are the two free parameters that get set by the phase transition temperature and relic dark matter abundance.

\subsection{Evolution prior to and at the phase transition to dark matter}
\label{SecEvol}

We define the dimensionless quantities
\begin{equation}\label{eDimLessDefs}
\Delta\equiv\frac{M}{T},\quad\kappa\equiv\frac{g_Y\varphi}{T},
\end{equation}
and the integral $\mathscr{I}_\alpha(x)\equiv\int_{x}^\infty dz\,\big(z^2-x^2\big)^\alpha/(e^z+1)$, in order to write the free energy density of the fermions as
\begin{equation}\label{eFFDefSimp}
    F_F=-\frac{2N_F}{3\pi^2\beta^4}\mathscr{I}_{\sfrac{3}{2}}(\kappa),
\end{equation}
following~\cite{Mandal:2019kkv}.
Noting that $d\mathscr{I}_{\sfrac{3}{2}}(\kappa)/d\kappa=-3\kappa\mathscr{I}_{\sfrac{1}{2}}(\kappa)$, Eq.~\eqref{eMassEq} becomes
\begin{equation}\label{eFJPotMassEq}
\frac{\pi^2\lambda}{2g_YN_F}\Delta^3=\kappa\,e^{\sfrac{\kappa\lambda}{g_Y\Delta}} \mathscr{I}_{\sfrac{1}{2}}(\kappa),
\end{equation}
folding in $U(\varphi)$ given in Eq.~\eqref{eFJPot}. Here $N_F$ is the number of fermion generations, which we take to be one in our model.
We discuss the evolution of this system below.

\begin{figure}[t]
    \begin{center}
    \includegraphics[width=\columnwidth]{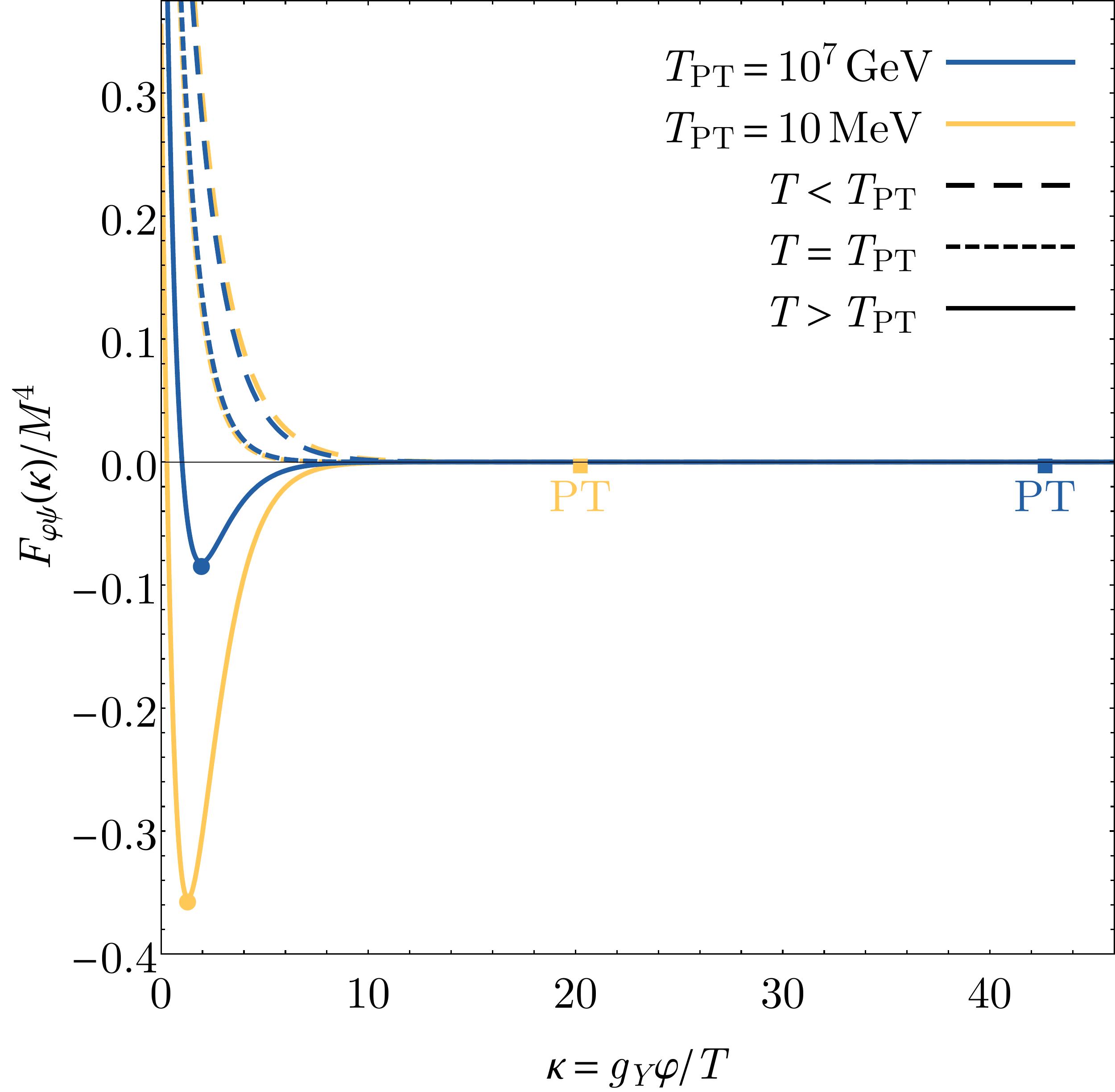}
    \end{center}
    \caption{Normalized free energy $F_{\varphi\psi}/M^4$, given by Eq.~\eqref{eTotalFreeEn}, as a function of $\kappa=g_Y\varphi/T$ for two different phase transition temperatures $T_{\rm PT}$. At $T>T_{\rm PT}$ the system is in a stable phase where $F_{\varphi\psi}$ has a minimum denoted by the filled circles. At the epoch of phase transition, $T=T_{\rm PT}$, the minimum is lost, and becomes an inflection point of the potential, as denoted by the filled squares. After the phase transition, the potential $F_{\varphi\psi}$ has no minimum, and the scalar field slowly rolls towards the global minimum of 0 at $\varphi=\infty$.}
    \label{FigNormalizedFR}
\end{figure}

At very high temperatures above the phase transition temperature, $T_{\rm PT}$, the coupled fermion-scalar system is in a stable phase where Eq.~\eqref{eFJPotMassEq} has a solution $\varphi_m$, where the free-energy density $F_{\varphi\psi}$ has a minimum such that $F_{\varphi\psi}(\varphi_m)<F_{\varphi\psi}(\infty)$.  As the universe cools, the system enters a metastable phase where the solution $\varphi_m$ corresponds to a local, as opposed to a global, minimum, i.e., $F_{\varphi\psi}(\varphi_m)>F_{\varphi\psi}(\infty)$.
Finally, when the temperature reaches $T_{\rm PT}$, the solution of Eq.~\eqref{eFJPotMassEq} is given by $\varphi=\varphi_\mathrm{PT}$, which we will derive below.  The loss of the minimum of $F_{\varphi\psi}$ at the phase transition makes $m_\varphi = 0$ from Eq.~\eqref{eRenormMass}.
After the phase transition, the system transitions into a phase with no equilibrium solution of Eq.~\eqref{eFJPotMassEq}.  In this phase, $m^2_\varphi<0$, which is unphysical for a particle; therefore we interpret the field $\varphi$ as a uniform rolling scalar field after the phase transition.
Qualitatively in this phase, the field $\varphi$ slowly rolls towards the global minimum of $F_{\varphi\psi}=0$ at $\varphi=\infty$.
We will see in Sec.~\ref{SecAfterPT} that in this phase, the combined scalar-fermion system behaves like dark matter.
In Fig.~\ref{FigNormalizedFR}, we show how the normalized free energy density, $F_{\varphi\psi}/M^4$, has a minimum before the phase transition, which it then looses after the phase transition.

\begin{figure}[t]
    \begin{center}
    \includegraphics[width=\columnwidth]{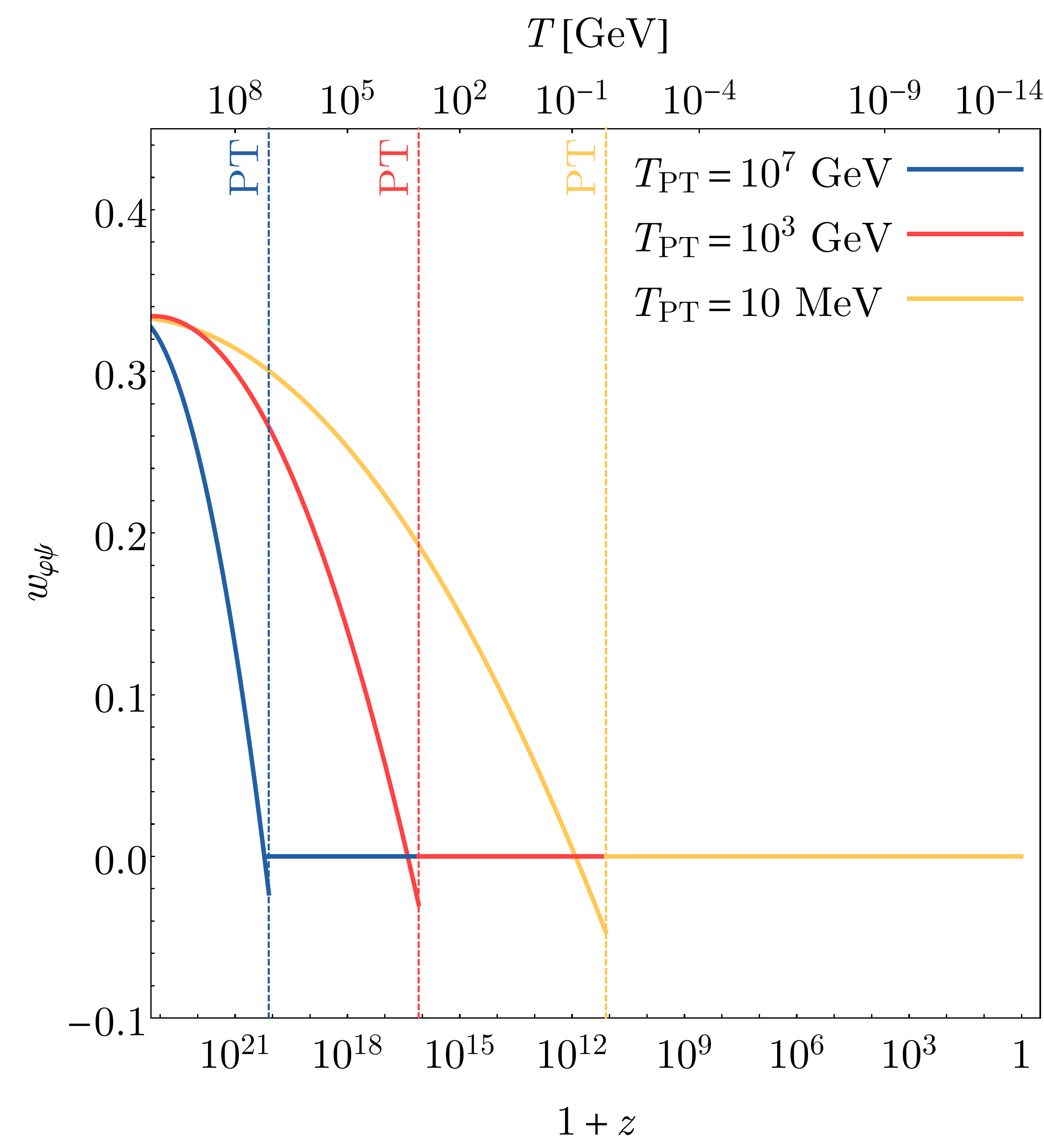}
    \end{center}
    \caption{Equation of state for the combined scalar-fermion system, $w_{\varphi\psi}$, as a function of redshift for several different phase transition temperatures $T_{\rm PT}$. The system initially behaves like radiation, and transitions to pressureless matter after the phase transition.}
    \label{FigEoS}
\end{figure}

The equation of state $w_{\varphi\psi}$ for the $\varphi\psi$ fluid is defined as $w_{\varphi\psi}\equiv P_{\varphi\psi}/\rho_{\varphi\psi}$, where $P_{\varphi\psi}=-F_{\varphi\psi}$ is the pressure of the combined system, and $\rho_{\varphi\psi}$ is its total energy density $\rho_{\varphi\psi}=U(\varphi)+2I_\epsilon(\kappa)/\pi^2\beta^4$, where
\begin{equation}\label{eScriDefs-3}
	I_\epsilon(x)\equiv\int_{x}^\infty\frac{dz\,z^2\big(z^2-x^2\big)^{\sfrac{1}{2}}}{e^z+1}.
\end{equation}
As shown in Fig.~\ref{FigEoS}, the system initially behaves like radiation with an equation of state parameter $w_{\varphi\psi}\approx 1/3$, and then changes after the phase transition to behave like dark matter with $w_{\varphi\psi}=0$.

\begin{figure}[t]
    \begin{center}
    \includegraphics[width=\columnwidth]{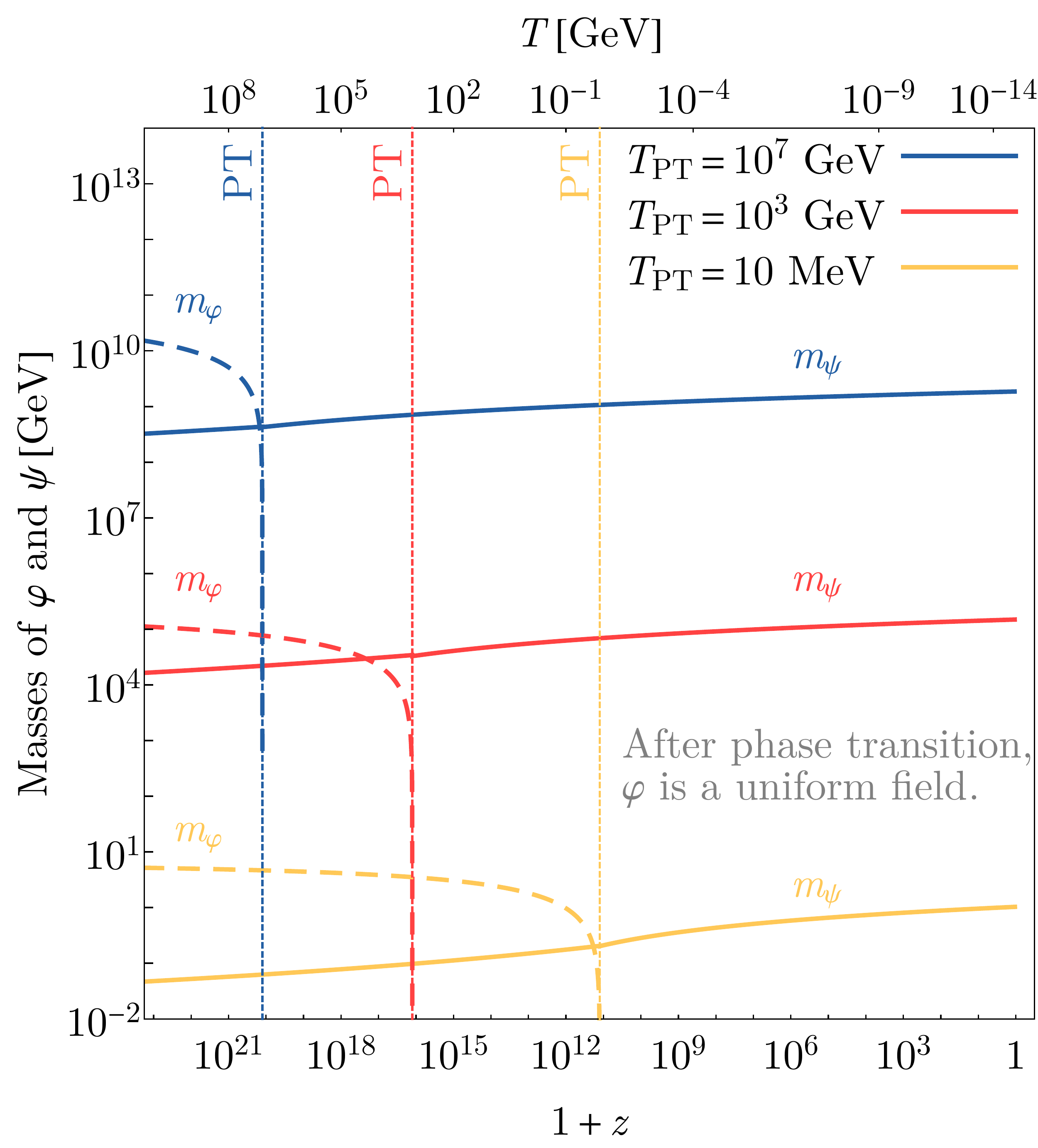}
    \end{center}
    \caption{Mass of the fermion (solid) and the scalar (dashed) from early times to today. The dotted vertical lines indicate the phase transition redshift for different choices of $T_{\rm PT}$. After the phase transition, the mass of the fermion slowly increases, and the scalar transitions from a particle to a uniform rolling field.}
    \label{FigMassesEvol}
\end{figure}

To solve for $\varphi_{\rm PT}$ at the phase transition, we make an analytic approximation of Eq.~\eqref{eFJPotMassEq}. We do this by using the fact that at $T=T_{\rm PT}$, the minimum of $F_{\varphi\psi}$ is lost, i.e. the second derivative becomes zero, and by assuming that the phase transition occurs in the regime where $g_Y\varphi\gg T$, which is valid given the current relic dark matter density.  This allows us to make the approximation $\mathscr{I}_{\sfrac{3}{2}}(\kappa)\approx 3\kappa^2K_2(\kappa)+\mathcal{O}(e^{-2\kappa})$, where $K_n(x)$ is the modified Bessel function of the second kind~\cite{BesselWolfram}.
Using Eq.~\eqref{eFJPotMassEq} yields the values at the phase transition of,
\begin{equation}\label{eDelCritKapCrit}
T_\mathrm{PT}=\frac{2g_Y M}{\lambda\left(\sqrt{1+\frac{Bg^{\sfrac{8}{3}}_Y}{\lambda^{8/3}}}+1\right)},\quad \varphi_\mathrm{PT}=\frac{3M}{\lambda\left(\sqrt{1+\frac{Bg^{\sfrac{8}{3}}_Y}{\lambda^{8/3}}}-1\right)},
\end{equation}
where $B=6\times2^{\sfrac{1}{3}}/\pi e$.
The redshift $z_{\rm PT}$ at which the phase transition occurs is obtained by noting that  $T_{\rm PT}$ can be related to the temperature today as $T_{\rm PT}=\left(1+z_{\rm PT}\right)T_0\left(g_{S,0}/g_{S,\mathrm{PT}}\right)^{\sfrac{1}{3}}$, where $g_S$ denotes the effective relativistic degrees of freedom.
This gives
\begin{equation}\label{eZCrit}
1+z_{\rm PT}=\frac{T_{\rm PT}}{T_0}\left(\frac{g_{S,\mathrm{PT}}}{g_{S,0}}\right)^{\sfrac{1}{3}}.
\end{equation}
The fermion number density $n_\psi$ is defined as $n_\psi=(1/4\pi^3)\int d^3p\,n_F(\epsilon)$, which holds at all times and does not assume that the fermion is continuously in equilibrium with the scalar\footnote{We note that at some time earlier than the phase transition point the fermion is in equilibrium with the scalar. For example, we calculate explicitly that for a fiducial reheating temperature of $10^{10}$ GeV and a Yukawa coupling constant $g_Y = 1$, we find that the two sectors are in equilibrium; specifically, the temperature-dependent rate of interaction $\Gamma_{\varphi\leftrightarrow \psi}$ between $\varphi$ and $\psi$ at $T=10^{10}\,{\rm GeV}$ is much larger the Hubble rate $H(T=10^{10}\,{\rm GeV})$ at that time.}. Since, in this Section, we are interested in the fermion number density at the phase transition temperature, we can use the limit $m_\psi=g_Y\varphi\gg T$, which we find later is a self-consistent assumption.  In this limit,
\begin{equation}\label{eFermCondDen2}
    n_\psi=\frac{1}{2}\frac{\pa F_F}{\pa m_\psi}=\frac{1}{2g_Y}\frac{\pa F_F}{\pa\varphi}=\frac{N_F m_\psi}{\pi^2\beta^2}\mathscr{I}_{\sfrac{1}{2}}(\kappa)
\end{equation}
using Eq.~\eqref{eFFDefSimp}, as was also calculated in~\cite{Chitov:2009ph}.
Solving for $\mathscr{I}_{\sfrac{1}{2}}(\kappa_{\rm PT})$ from Eq.~\eqref{eFJPotMassEq} yields
\begin{equation}\label{eRhoSCrit}
\frac{n_{\psi,\mathrm{PT}}}{M^3}=\frac{\lambda}{2g_Y} e^{-{\sfrac{\lambda\varphi_{\rm PT}}{M}}}=\frac{\lambda}{2g_Y} \exp\left(-\frac{3}{\sqrt{1+\frac{B g^{\sfrac{8}{3}}_Y}{\lambda^{8/3}}}-1}\right).
\end{equation}
which gives the number density of fermions at the phase transition. In Fig.~\ref{FigMassesEvol} we show the masses of the fermion $\psi$ and the scalar $\varphi$ as a function of redshift $z$.
The scalar mass asymptotically approaches zero at the phase transition, after which it becomes a uniform rolling field without a particle interpretation.

The phase transition described above is a first-order phase transition.
This can be seen because as the temperature of the $\varphi\psi$ system approaches $T_{\rm PT}$, the fermion mass approaches its critical value $m_{\psi,\mathrm{PT}}$ as $m_\psi-m_{\psi,\mathrm{PT}}\approx(T-T_\mathrm{PT})^{\sfrac{1}{2}}$.
This critical exponent of $\frac{1}{2}$ is a signature of a first-order phase transition~\cite{Quiros:1999jp, Hindmarsh:2020hop, Peskin:1995ev, Goldenfeld:1992xx}.  While sometimes first-order phase transitions are characterized by a barrier between the true and false vacua, which remains even after the phase transition, in our model the false vacuum and the barrier disappear at $T=T_{\rm PT}$.
The system then starts rolling towards the true vacuum at $\varphi=\infty$.
Since the true vacuum is at $\infty$, rather than at a finite value of $\varphi$, the transition of the field from the false to the true vacuum is not instantaneous.
However, we have confirmed that the energy density of the system undergoes an abrupt change at the phase transition point by calculating $\rho_{\varphi\psi}$ before the phase transition, given above Eq.~\eqref{eScriDefs-3}, and $\rho_{\psi} + \rho_{\varphi}$ after the phase transition, given by Eq.~\eqref{eDMEnDen}.  Note, however, that $\rho_{\varphi}$ is continuous through the phase transition.
Moreover, there is a discontinuity in the first derivative of the free energy with respect to the scalar field, as seen from Eq.~\eqref{eSaddleConditions}, where the derivative is always zero before the phase transition and non-zero after; this is also indicative of a first-order phase transition.

\subsection{Evolution after the phase transition}
\label{SecAfterPT}

\begin{figure}[t]
    \begin{center}
    \includegraphics[width=\columnwidth]{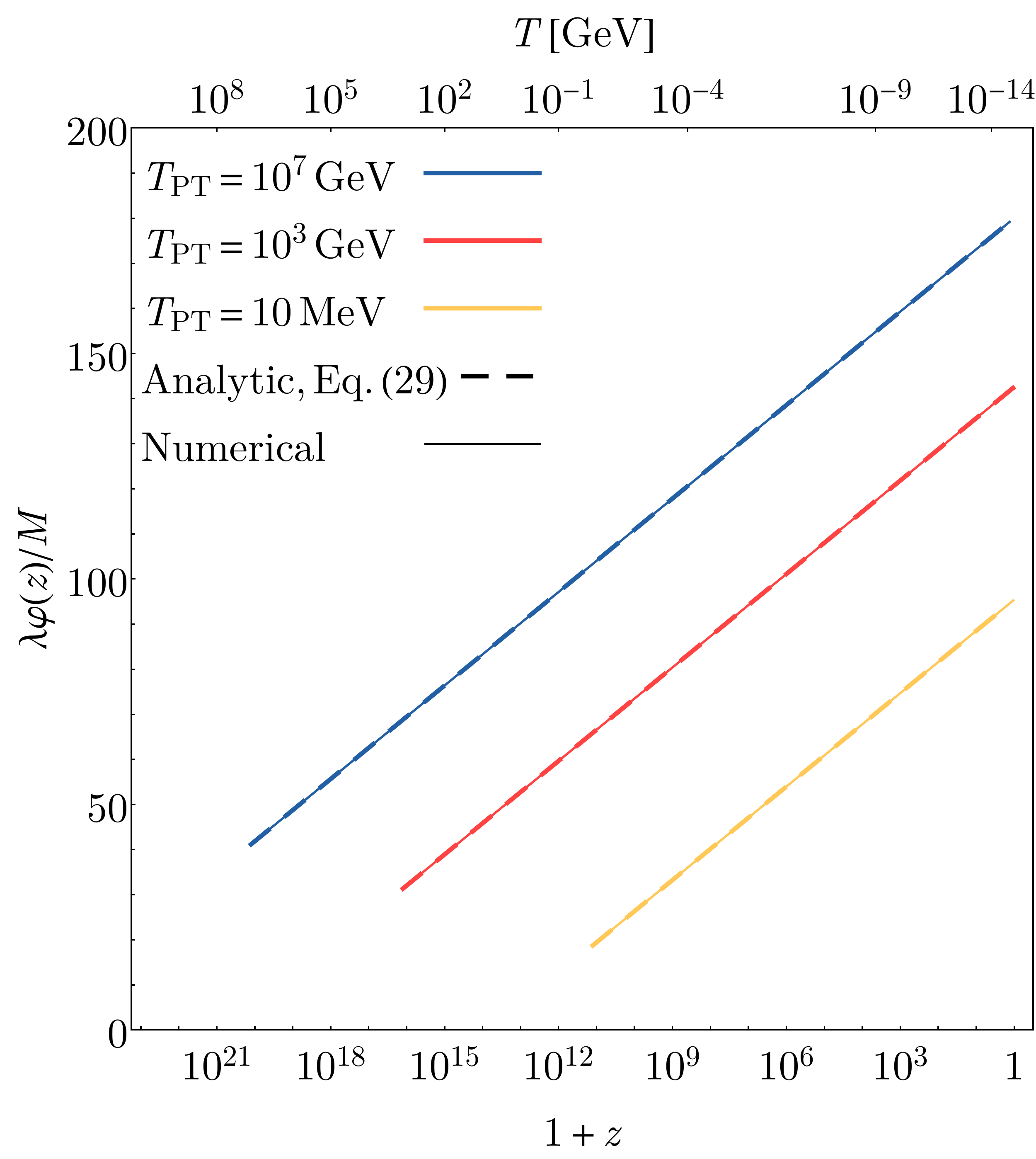}
    \end{center}
    \caption{Comparison of the exact numerical solution $\varphi$ of Eq.~\eqref{eScalarEvolAfter} with the analytic solution in Eq.~\eqref{eSolDMPhase}, as a function of redshift for several choices of the phase transition temperature $T_{\rm PT}$; this shows the validity of Eq.~\eqref{eSolDMPhase} as an accurate approximation to the true numerical solution.  Here we scale $\varphi$ by $\lambda/M$ for visual purposes.}
    \label{FigNumSolZ}
\end{figure}

As mentioned above, in the dark-matter phase Eq.~\eqref{eSaddleConditions} no longer holds, and one needs to solve the full equation for $\varphi$ given by
\begin{equation}\label{eScalarEvolFull}
	\ddot{\varphi}+3H\dot{\varphi}+\frac{\pa F_{\varphi\psi}}{\pa\varphi}=0,
\end{equation}
as opposed to using Eq.~\eqref{eFJPotMassEq}.  Here the dot denotes derivatives with respect to $t$, and $H$ is the Hubble rate determined from the Friedmann equation
\begin{equation}\label{eFriedmannEqs}
H^2(t)=\left(\frac{\dot{a}}{a}\right)^2=\frac{8\pi G}{3}\rho_\mathrm{tot},
\end{equation}
where $\rho_\mathrm{tot}$ is the total energy density of the Universe.  Analogous to Eq.~\eqref{eFermMass}, we define the fermion mass in this phase to be $g_Y$ times the solution ${\varphi}(t)$ to Eq.~\eqref{eScalarEvolFull}, i.e.,
\begin{equation}\label{eFermMassDMPhase}
m_\psi(z)\equiv g_Y\varphi(z).
\end{equation}
Since
\begin{equation}
    \frac{\pa F_{\varphi\psi}}{\pa\varphi}=\frac{\partial U(\varphi)}{\partial\varphi}+\frac{\pa F_F}{\pa \varphi}
\end{equation}
and $\pa F_F/\pa\varphi=2g_Y n_\psi$ from Eq.~\eqref{eFermCondDen2}, then
\begin{equation}\label{eScalarEvolAfter}
	\ddot{\varphi}+3H\dot{\varphi}=-\frac{\pa U}{\pa\varphi}-2g_Y n_{\psi,\mathrm{PT}}\left(\frac{a_{\rm PT}}{a}\right)^3.
\end{equation}
Here we note that $n_\psi \propto a^{-3}$ in the dark matter phase since $\psi$ is already a heavy particle at the phase transition point.

\begin{figure}[t]
    \begin{center}
    \includegraphics[width=\columnwidth, height=9.8cm]{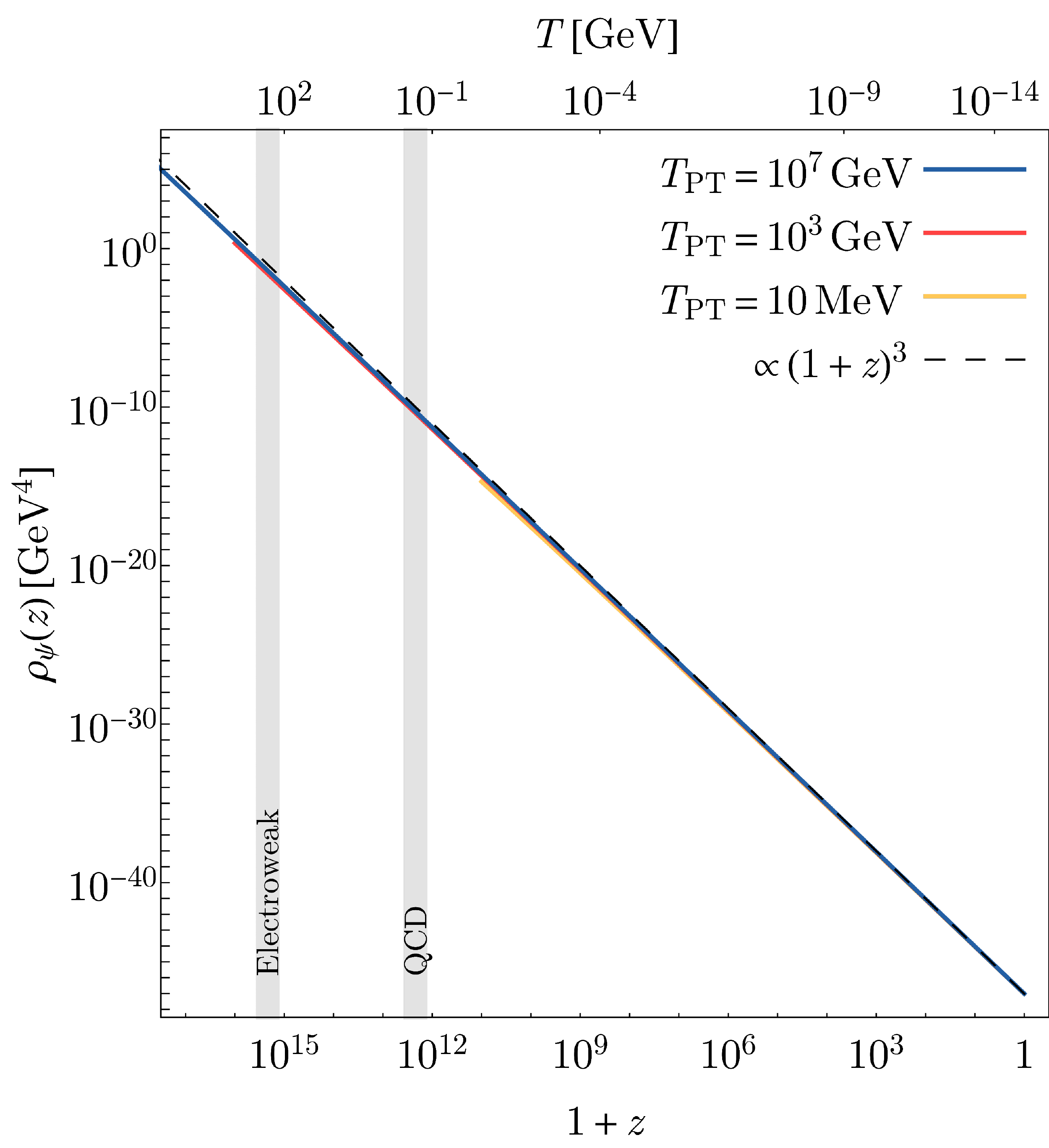}
    \end{center}
    \caption{Energy density of the fermion component, $\rho_\psi$, as a function of redshift for several choices of the phase transition temperature $T_{\rm PT}$. The dashed black line shows an energy density component diluting strictly as $(1+z)^3$, and the variation of the fermion energy density in Eq~\eqref{ePsiEnDen} gives a small deviation from this.}
    \label{FigPsiEnDen}
\end{figure}

We obtain the solution $\varphi$ in the dark matter phase by solving Eq.~\eqref{eScalarEvolAfter} numerically, and we show this solution in Fig.~\ref{FigNumSolZ}.
We can also obtain an approximate analytic solution by guessing that the terms on the left-hand side of Eq.~\eqref{eScalarEvolAfter} are negligibly small compared to the terms on the right-hand side.  
This gives, using Eq.~\eqref{eRhoSCrit},
\begin{equation}\label{eMEstimate}
\frac{\pa U}{\pa \varphi}=-\lambda M^3\exp\left(-\frac{3}{\sqrt{1+\frac{B g^{\sfrac{8}{3}}_Y}{\lambda^{8/3}}}-1}\right)\left(\frac{a_{\rm PT}}{a}\right)^3.
\end{equation}
Since we also have
\begin{equation}\label{eUPhiDeriv}
\frac{\pa U}{\pa \varphi}=-\lambda M^3 e^{-{\sfrac{\lambda\varphi}{M}}},
\end{equation}
then we can equate the right-hand sides of Eqs.~\eqref{eRhoSCrit} and ~\eqref{eMEstimate}.
Taking the natural logarithm of both sides gives us the explicit form of the analytic solution ${\varphi}$,
\begin{equation}\label{eSolDMPhase}
    {\varphi}(z)=\frac{3M}{\lambda}\left[\frac{1}{\sqrt{1+\frac{B g^{\sfrac{8}{3}}_Y}{\lambda^{8/3}}}-1}+\ln\frac{1+z_{\rm PT}}{1+z}\right].
\end{equation}
We show in Fig.~\ref{FigNumSolZ} that this analytic solution is a good match to the exact numerical solution.
As a cross check, when substituting the analytic solution into Eq.~\eqref{eScalarEvolAfter}, we find that the two terms on the right-hand side are many orders of magnitude larger than the terms on the left-hand side, effectively making the latter negligible.  We note however, that this is the case only for coupling constants $g_Y$ larger than $10^{-10}$; this is because for $g_Y$ smaller than this, the solution to Eq.~\eqref{eScalarEvolAfter} no longer leads to dark matter-like behavior. For $g_Y$ up to 1, we find the numerical solution matches the analytic solution for phase transition temperatures up to $10^7$~GeV; formally, for $g_Y$ values up to $10^{10}$ the analytic and numeric solutions match well for phase transition temperatures up to $10^{17}$~GeV, which would yield dark matter masses at the Planck scale, but we restrict to $g_Y\leq 1$ and an upper bound of $10^7$~GeV in this work.
We bound the phase-transition temperatures below to be above 10 MeV to ensure the dark matter is in place prior to Big Bang nucleosynthesis.

From Eq.~\eqref{eSolDMPhase} we see that the fermion mass increases weakly with redshift as shown in Fig.~\ref{FigMassesEvol}, however, $\rho_\psi(z)$ is still well approximated by an $a^{-3}$ dependence as we show in Fig.~\ref{FigPsiEnDen}.
Explicitly, the fermion mass at the present time is given by
\begin{equation}\label{eMassPsiNow}
m_{\psi,0}=\frac{3g_YM}{\lambda}\left[\frac{1}{\sqrt{1+\frac{B g^{\sfrac{8}{3}}_Y}{\lambda^{8/3}}}-1}+\ln(1+z_{\rm PT})\right].
\end{equation}
The energy density of the fermion is given by
\begin{equation}\label{ePsiEnDen}
\begin{aligned}
    \rho_\psi(z)&=m_\psi(z)n_\psi(z)\\
    &=\frac{3M^4}{2}\left[\frac{1}{\sqrt{1+\frac{B g^{\sfrac{8}{3}}_Y}{\lambda^{8/3}}}-1}+\ln\frac{1+z_{\rm PT}}{1+z}\right]\\
    &\times\exp\left(-\frac{3}{\sqrt{1+\frac{B g^{\sfrac{8}{3}}_Y}{\lambda^{8/3}}}-1}\right)
    \left(\frac{1+z}{1+z_{\rm PT}}\right)^3.
\end{aligned}
\end{equation}
The above discussion demonstrates that both the scalar and the fermion components behave as pressureless dark matter after the phase transition.

Since 
\begin{equation}
    \frac{\pa U}{\pa\varphi}\propto U(\varphi) \text{ and }\rho_{\varphi}=U(\varphi),
\end{equation}
then
\begin{equation}\label{ePhiEnDenDef}
\rho_{{\varphi}}(z)=M^4 \exp\left(-\frac{3}{\sqrt{1+\frac{B g^{\sfrac{8}{3}}_Y}{\lambda^{8/3}}}-1}\right)
    \left(\frac{1+z}{1+z_{\rm PT}}\right)^3
\end{equation}
from Eq.~\eqref{eMEstimate}.
Thus, the energy density $\rho_{{\varphi}}$ has an $a^{-3}$ dependence, which makes it behave as pressureless matter after the phase transition. Eq.~\eqref{eSolDMPhase} shows that after the phase transition, ${\varphi}$ slowly increases to $\infty$.

\section{Phenomenological implications}
\label{secPheno}

Our model effectively has only one free parameter, which is the phase transition temperature, $T_{\rm PT}$, once the relic density of dark matter today,  $\Omega_{\rm DM,0}$, is set to the observed value.
We fix $\Omega_{\rm DM,0} =\rho_{\rm DM,0}/\rho_{\rm tot,0}$ to 0.26 in order to match the latest \textit{Planck} 2018 TT, TE, EE + lowE + lensing + BOSS BAO data~\cite{Planck:2018vyg}, which then determines $M$ and $\lambda$.

The total dark matter energy density, $\rho_{\rm DM}$, is a sum of the energy densities of both ${\varphi}$ and $\psi$, which is given by
\begin{equation}\label{eDMEnDen}
\begin{aligned}
\rho_{\rm DM}(z)&=M^4\left\{1+\frac{3}{2}\left[\frac{1}{\sqrt{1+\frac{B g^{\sfrac{8}{3}}_Y}{\lambda^{8/3}}}-1}+\ln\frac{1+z_{\rm PT}}{1+z}\right]\right\}\\
    &\times\exp\left(-\frac{3}{\sqrt{1+\frac{B g^{\sfrac{8}{3}}_Y}{\lambda^{8/3}}}-1}\right)
    \left(\frac{1+z}{1+z_{\rm PT}}\right)^3.
\end{aligned}    
\end{equation}
from Eqs.~\eqref{ePsiEnDen} and~\eqref{ePhiEnDenDef}.  Using Eq.~\eqref{eZCrit} and~\eqref{eMassPsiNow} gives the present abundance of dark matter in terms of the phase transition temperature as
\begin{equation}\label{eRhoDMNow}
\frac{\rho_{\rm DM,0}T^3_{\rm PT}g_{S,\rm PT}}{M^4 T^3_{0}g_{S,0}}=\exp\left(-\frac{3}{\sqrt{1+\frac{B g^{\sfrac{8}{3}}_Y}{\lambda^{8/3}}}-1}\right)\left\{1+\frac{\lambda m_{\psi,0}}{2g_Y M}\right\}.
\end{equation}
For the above, we will assume $g_{S,0}=3.91$ and $g_{S,\mathrm{PT}}=100$ as the effective degrees of freedom.  We note that changing $g_{S,\mathrm{PT}}$ by an order of magnitude, either higher or lower, only changes $m_{\psi,0}$ by a few percent.

\begin{table}[t]
	\centering
	\begin{tabular}{ccccc}
		\hline\hline
        $\bm{T_{\rm PT}}$ & & $~~~\bm{m_{{\rm{DM}},0}\,{\rm [GeV]}}$ & &~~~$\bm{\Omega_{{{\rm{DM}},0}}^{\rm{CMB}}/\Omega_{{{\rm{DM}},0}}^{\rm{Late-time}}}$ \\
		\hline
		$10^7$~GeV & & $2.0\times 10^9$ & & $0.88$ \\
		$10^3$~GeV & & $1.5\times 10^5$ & & $0.84$ \\
		 10~MeV & & $1.0$ & & $0.78$ \\
		\hline\hline
	\end{tabular}
	\caption{We show for a few specific values of the phase transition temperature, $T_{\rm PT}$, the present mass $m_{{\rm{DM}},0}$ of the dark matter, given by $m_{\psi,0}$, and the fractional discrepancy in the relic energy density of the dark matter, ${\Omega_{{\rm{DM}},0}}$, inferred from measurements at recombination versus late-time measurements assuming $\Lambda$CDM. The latter discrepancy arises from the variation of the fermion energy density as $\ln(1/1+z)$ shown in Eq.~\eqref{ePsiEnDen}, which results in early-time inferences of ${\Omega_{{\rm{DM}},0}}$ being slightly smaller than late-time measurements if a $(1+z)^{3}$ scaling of $\rho_{{\rm{DM}}}$ is assumed.
 	Here we set $\Omega_{\rm{DM},0}^{\rm Late-time}=0.26$ to match the {\it{Planck}} primary CMB+{\it{Planck}} lensing+BOSS BAO data~\cite{Planck:2018vyg}.
	}
	\label{TabFJMass}
\end{table}

\begin{figure}[t]
    \begin{center}
    \includegraphics[width=1.0\columnwidth]{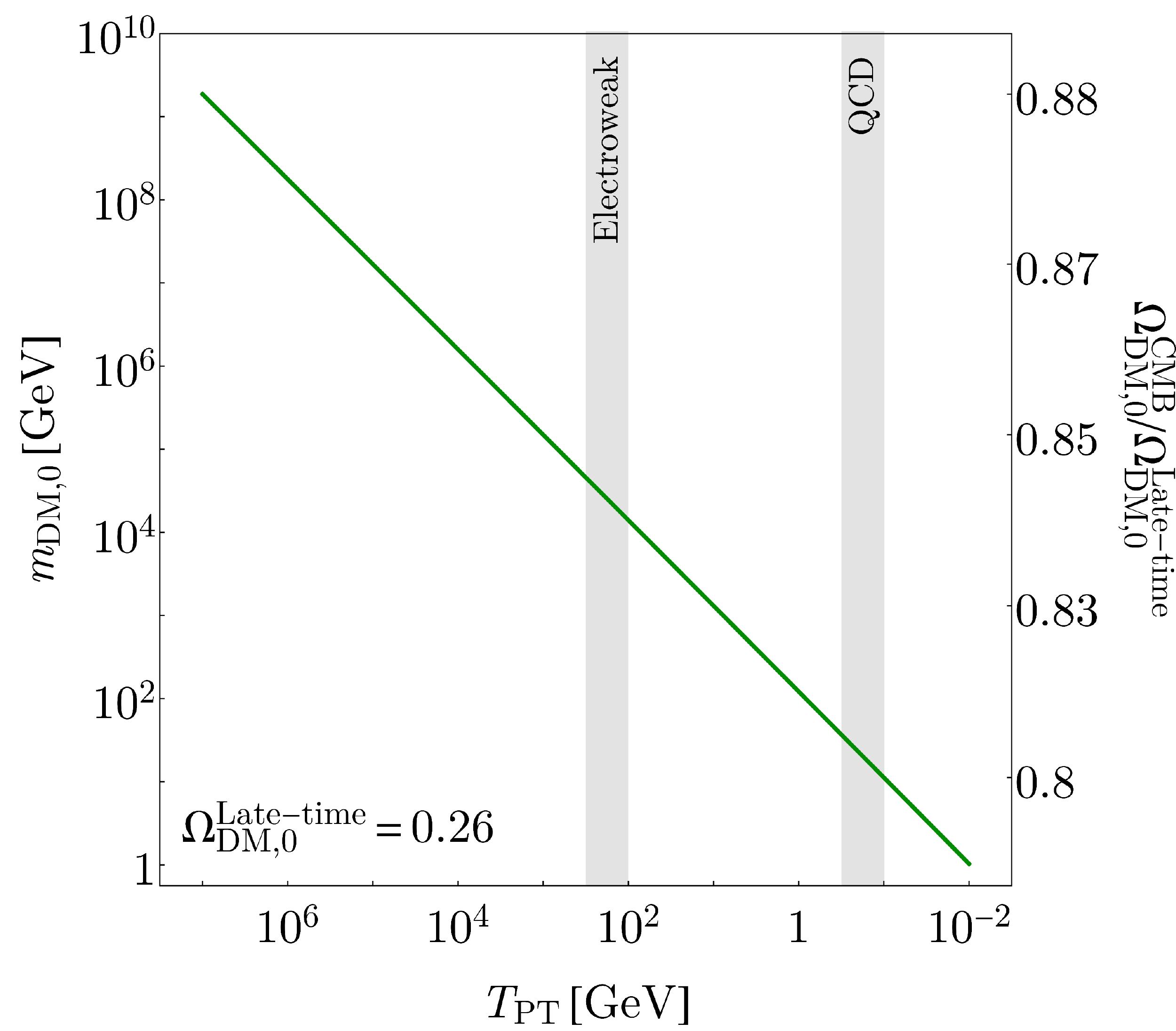}
    \end{center}
    \caption{Mass of the dark matter at the present time, $m_{\rm DM,0}$, as a function of the phase transition temperature $T_{\rm PT}$. Here the relic dark matter energy density $\Omega_{\rm DM,0}^{\rm Late-time}$ has been fixed to 0.26 to match the {\it{Planck}} primary CMB+{\it{Planck}} lensing+BOSS BAO data~\cite{Planck:2018vyg}. We find masses in a range of interest for a dark matter particle. We also show the fractional discrepancy in the relic energy density of the dark matter, ${\Omega_{{\rm{DM}},0}}$, inferred from measurements at recombination versus late-time measurements assuming $\Lambda$CDM; given a late-time measurement of ${\Omega_{{\rm{DM}},0}}$, this model predicts a lower amount of dark matter at early times than predicted by $\Lambda$CDM.}
    \label{FigMassesNow}
\end{figure}

\begin{figure}[t]
    \begin{center}
    \includegraphics[width=1.0\columnwidth]{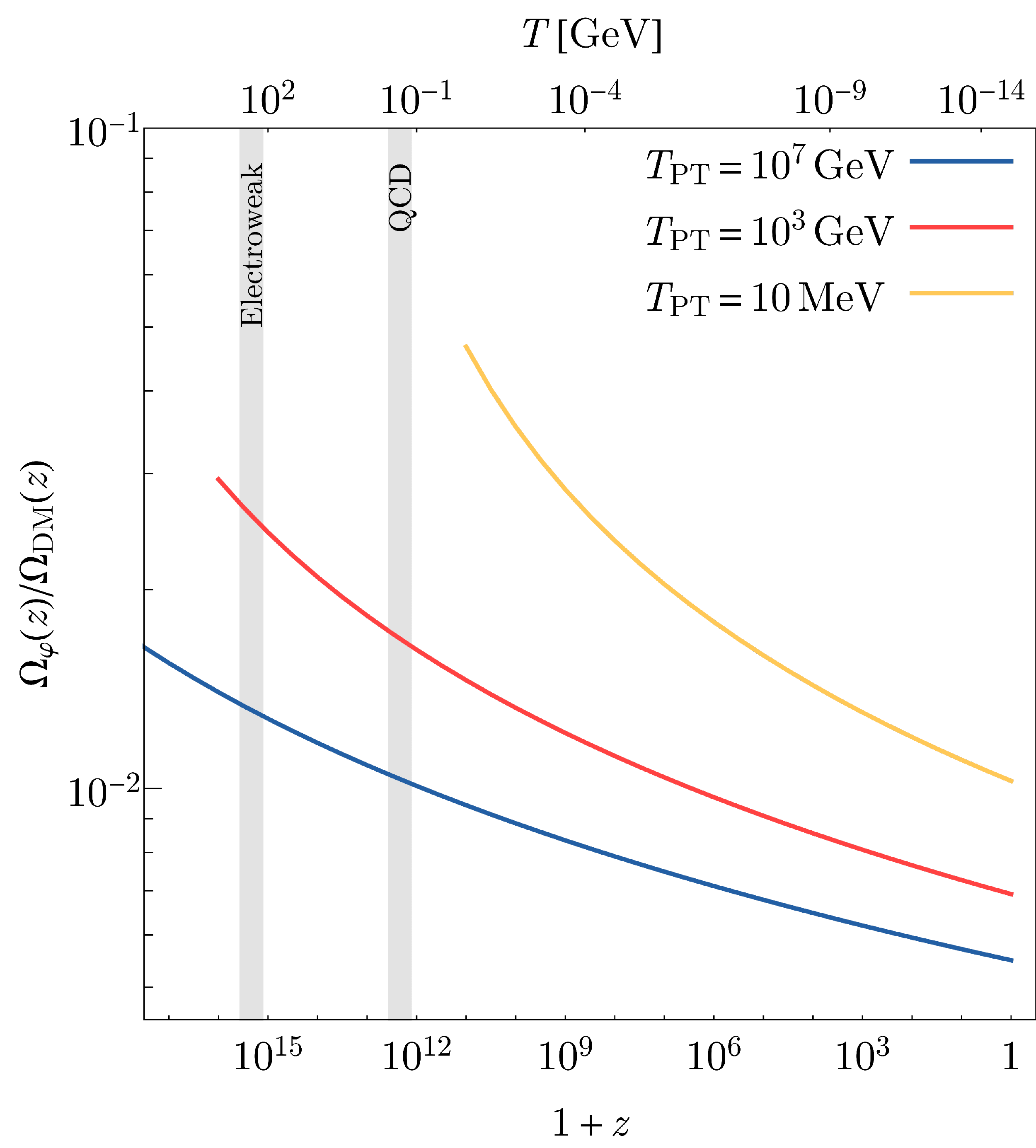}
    \end{center}
    \caption{Fraction of the dark matter energy density contained in the unclustered scalar field, $\varphi$, as a function of redshift for several different phase transition temperatures $T_{\rm PT}$.  We see the scalar field is a sub-dominant component of the dark matter that can range from a few to sub-percent of the total dark matter energy density after the phase transition.}
    \label{FigOmPhiZ}
\end{figure}

In Tab.~\ref{TabFJMass} and Fig.~\ref{FigMassesNow}, we show the mass of the fermion $\psi$ at the present time as a function of phase-transition temperature $T_{\rm PT}$.  
We find that for $T_{\rm PT}$ in the range of 10 MeV to $10^7$~GeV, we obtain $m_{{\rm{DM}},0}$, given by $m_{\psi,0}$, in the range of 1 GeV to $2 \times 10^9$ GeV.  We also see that the $\ln(1/1+z)$ dependence of $\rho_\psi(z)$ in Eq.~\eqref{ePsiEnDen} leads to a slight deviation of $\rho_\psi(z)$ away from a strict $(1+z)^3$ scaling.  We note that while $m_{{\rm{DM}},0}$ is calculated here assuming a flat Universe, the $\ln(1/1+z)$ dependence is independent of the flatness assumption. This deviation results in a discrepancy in the energy density of the dark matter today ${\Omega_{{\rm{DM}},0}}$ inferred from measurements at recombination versus late-time measurements assuming a $\Lambda$CDM scaling.  As shown in Tab.~\ref{TabFJMass} and Fig.~\ref{FigMassesNow}, the discrepancy between the two quantities can range from $10\%$ to $20\%$ for the allowed range of phase-transition temperatures. 
In Fig.~\ref{FigOmPhiZ}, we show the fraction of the dark matter energy density comprised by the unclustered scalar field, i.e.~$\Omega_\varphi/\Omega_{\rm DM}\equiv\rho_\varphi/\rho_{\rm DM}$, as a function of redshift for several choices of the phase transition temperature $T_{\rm PT}$. We see that the scalar field is a sub-dominant component of the dark matter that can range from a few to sub-percent of the total dark matter density after the phase transition.

\section{Discussion}
\label{secDiscuss}

We have proposed a dark matter model comprised of a scalar field, with an exponential self-interaction potential, interacting via a Yukawa coupling with a fermionic field. 
Studying the evolution of this system, we find that as the Universe cools, the coupled system undergoes a phase transition after which it behaves like pressureless dark matter; in this dark matter phase, the mass of the heavy fermion increases over time and the scalar acts as a uniform, unclustered rolling field. This model contains only the phase-transition temperature as a free parameter once the relic dark matter abundance today is matched to observations.

The novelty of this model lies in the fact that both the mass of the fermion and the dark matter-like behavior of the system at late times arises from the temperature-dependent interaction between the fermion and the scalar.
The choice of the potential for the scalar field, which might arise naturally from a higher energy theory involving supersymmerty, supergravity, or compactified extra dimensions~\cite{Ferreira:1997hj}, enables this unique behavior.
The mass of the fermion lies in the range of 1~GeV to $10^9$~GeV for allowed phase-transition temperatures in the range 10~MeV to $10^7$~GeV, which are mass and temperature ranges of interest for dark matter.

The cosmological observational signatures of this model are that the unclustered scalar field leads to a  scale-independent suppression of structure formation of order a percent.
This would lead to a discrepancy between the amplitude of structure formation derived from expansion rate probes in the context of $\Lambda$CDM, versus direct structure growth measurements.
A larger observable feature arises from the mass-varying fermionic component, which results in a discrepancy of about ten percent between the dark matter relic density, ${\Omega_{{\rm{DM}},0}}$, measured today and that inferred from measurements at recombination assuming a $(1+z)^{3}$ scaling of $\rho_{{\rm{DM}}}$ to extrapolate it to today.  Such behavior may help to alleviate some cosmological tensions recently discussed in the literature~\cite{Verde:2019ivm,Freedman:2017yms,Planck:2018vyg,Riess:2021jrx,DiValentino:2020vnx,DiValentino:2020zio, Pesce:2020xfe,Addison:2017fdm, Freedman:2019jwv,DES:2017myr,Hildebrandt:2018yau,HSC:2018mrq, KiDS:2020suj,DES:2021wwk,Nunes:2021ipq}.
In general, both the $H_0$ and $\sigma_8$ tensions can be alleviated if the amount of dark matter between recombination and today is lower than predicted by $\Lambda$CDM; a lower amount of dark matter would suppress structure growth, and would lead to an earlier onset of dark energy domination increasing low-redshift $H_0$ measurements. Whether the model presented here yields less or more dark matter than predicted by $\Lambda$CDM depends on the redshift at which direct measurements of the dark matter density are being made; for direct low-redshift measurements of the dark matter density, this model predicts less dark matter at all earlier times than predicted by $\Lambda$CDM.  However, for direct measurements of the dark matter density at recombination, this model predicts more dark matter at all later times than predicted by $\Lambda$CDM.  We note, however, that primordial CMB data only constrains the density of dark matter at recombination under the assumption of flatness and perfect knowledge of the sum of the masses of the neutrinos.  Relaxing flatness and neutrino mass assumptions, CMB inferences of the dark matter density are well constrained only by adding low-redshift observations, for example from CMB lensing or BAO.   We leave further investigation in this direction to future work.

This simple model can be extended by coupling the fermion and scalar fields weakly to the standard model beyond what we have focused on this analysis. For example, as mentioned previously, the scalar field will couple to the Higgs field, which makes the fields of our model couple with standard model particles non-gravitationally.  Additionally, more complex interactions between the dark sector and the standard model are possible.
In such cases, this dark sector could be probed via direct-detection experiments which involve collisions of the dark matter particles with standard model particles, via indirect-detection experiments which look for the standard model products of dark matter self-annihilations or decays, and/or via particle colliders. One can also extend the dark sector component of the model further to construct more complex models with both the scalar and the fermionic sectors comprising multiple fields, and an appropriately generalized Yukawa interaction among those fields. In addition, the fermions can be charged under some internal gauge symmetry of the dark sector.  The phenomenological implications of such an extended dark sector might lead to additional interesting observational signatures.

\begin{acknowledgments}
We thank Kim Berghaus, Peizhi Du, Rouven Essig, Amanda MacInnis, Patrick Meade, Chaitanya Prasad, Anton Rebhan, Martin Ro\v{c}ek, Mauro Valli, and Peter van Nieuwenhuizen for useful discussions.  We also thank the anonymous referee for their careful reading and comments that improved the draft. SM and NS acknowledge support from NSF Grant number AST-1907657 and DOE Award number DE-SC0020441.
SM also acknowledges support from the Shota Rustaveli National Science Foundation (SRNSF) of Georgia (grant FR/18-1462).
\end{acknowledgments}

\appendix

\section{Vacuum quantum corrections}
\label{Zero-temp}

In our work the scalar potential $U(\varphi)$ is given by Eq.~\eqref{eFJPot}, and its exponential form leads to the solution of Eq.~\eqref{eScalarEvolAfter}
given by Eq.~\eqref{eSolDMPhase}.
This solution ensures that the mass $m_\psi$ of the fermion only has a weak logarithmic dependence with redshift in addition to the expected $(1+z)^3$ dependence so that its energy density does not deviate significantly from the $\Lambda$CDM expectation; see discussion in Sec.~\ref{secPheno}.
In addition, it also ensures the $(1+z)^3$ dependence of the energy density of the scalar component.

We assume in this work that the scalar potential $U(\varphi)$ of the form in Eq.~\eqref{eFJPot} arises from a higher energy theory.  Such exponential potentials of scalar fields arise generically in several classes of fundamental theories at high energies~\cite{Ferreira:1997hj}.
For example, Kaluza-Klein theories contain extra dimensions which, when compactified to a four-dimensional theory, produce an effective scalar field with an exponential potential~\cite{Wetterich:1984wd,Halliwell:1985bx}.
Exponential potentials also arise in superstring and supergravity theories~\cite{Cremmer:1983bf, Ellis:1983sf, Nishino:1984gk, Witten:1985xb, Dine:1985rz}, and in higher-order gravity theories~\cite{Barrow:1988xh}.
We also assume in this work that adding a coupling of the scalar with a fermion, given by the Yukawa coupling in Eq.~\eqref{eMainLag} does not destroy the higher energy fundamental physics generating the exponential potential, for example, given that the coupling is not supersymmetric.

In addition, when adding this coupling between the scalar and fermion, a significant zero-temperature quantum correction to the potential $U(\varphi)$ arises unless $U(\varphi)$ is considered to be an effective low-energy potential with the vacuum quantum correction included. This zero-temperature quantum correction arises from the fermionic one-loop contribution to the potential obtained by summing all one-particle irreducible diagrams\footnote{A one-particle irreducible  diagram is one which cannot be split into two nontrivial diagrams by cutting a single line anywhere in the diagram.} with one fermionic loop to get
\begin{equation}\label{e1Loop-1}
    V_{\rm 1-loop}^{\rm fermion}(\varphi)=-2\int\frac{d^4p}{(2\pi)^4}\ln\left[p^2+m^2_\psi(\varphi)\right],
\end{equation}
where $m_\psi(\varphi)=g_Y\varphi$ as in Eq.~\eqref{eFermMassDMPhase}.
Since this integral diverges as $p\rightarrow\infty$, we regularize it by introducing a cutoff $\Lambda$; we integrate from $p=0$ to $\Lambda$ and then remove the $\Lambda$-dependence from physical results by imposing the renormalization condition that the contribution $V^{\rm fermion}_{\rm 1-loop}(\varphi_{\rm PT})$ is zero at the phase transition point.  This gives
\begin{equation}\label{e1Loop-5}
    V^{\rm fermion}_{\rm 1-loop}(\varphi)=-\frac{g^4_Y\varphi^4}{16\pi^2}\ln\frac{\varphi^2}{\varphi^2_{\rm PT}}.
\end{equation}
There is also a bosonic one-loop contribution to the potential, given by
\begin{equation}\label{e1LoopBos}
    \begin{aligned}
    V^{\rm boson}_{\rm 1-loop}(\varphi)&=\frac{g^2_Y}{64\pi^2}\left(\frac{\pa^2F_{\varphi\psi}}{\pa\varphi^2}\right)^2\ln\left(\frac{1}{g^2_Y\varphi^2}\frac{\pa^2F_{\varphi\psi}}{\pa\varphi^2}\right).
    \end{aligned}
\end{equation}
Here $F_{\varphi\psi}$ is the tree-level free-energy density given in Eq.~\eqref{eTotalFreeEn}.
We can ignore this bosonic one-loop contribution since it takes on complex values away from the minimum of the potential where $\pa^2F_{\varphi\psi}/\pa\varphi^2<0$.

The fermionic correction $V^{\rm fermion}_{\rm 1-loop}(\varphi)$ is around $\mathcal{O}(10^{40}-10^{80})$ larger than $U(\varphi)$ today, depending on $T_{\rm PT}$, and thus it would naively destroy the shape of $F_{\varphi\psi}$ far from the phase transition.  In the dynamics of our model, $\pa F_{\varphi\psi}/\pa\varphi$ is the important quantity, and we should compare $\pa V^{\rm fermion}_{\rm 1-loop}(\varphi)/\pa\varphi$ to $\pa U(\varphi)/\pa\varphi$ to determine the relative importance of the vacuum correction.  We note that the former is $\propto g^4_Y/\lambda^3$ using Eq.~\eqref{eSolDMPhase} for $\varphi$, and the latter is $\propto \lambda$ from Eq.~\eqref{eUPhiDeriv}.  Thus, as we discussed in Section~\ref{secPot}, since $g^4_Y/\lambda^4$ is always fixed in our model, the relative importance of $V^{\rm fermion}_{\rm 1-loop}(\varphi)$ cannot be diluted by lowering the coupling constant $g_Y$, as may naively be assumed.

A solution is to view the Lagrangian given by Eq~\eqref{eMainLag} as an effective low-energy description that includes the zero-temperature fermionic quantum correction.  We consider it premature to justify our Lagrangian on the basis of a higher energy UV-complete particle physics theory at this point, and further investigation is warranted on how specifically it may arise as a low-energy effective description of a class of theories based on the framework of supersymmetry, superstring theory, higher-order gravity theories, or Kaluza-Klein theories.  In this work, we only demonstrate that given this Lagrangian we find a viable model of dark matter that fits observational constraints. 
We note that a similar situation of suppressing zero-temperature quantum corrections may arise for possible descriptions of inflation where quantum corrections arise from the coupling of the scalar inflaton field with matter, as is expected for reheating to occur; such vacuum quantum corrections can potentially destroy the shape of the inflaton potential unless suppressed by a higher energy mechanism. We leave further investigation of these issues to future work.

\section{Coupling of the scalar with the Higgs}
\label{AddScalarInt}

The scalar potential $U(\varphi)$ will also receive corrections from the interaction of $\varphi$ with other fields present in the Universe, consistent with the underlying symmetries governing these fields.
Of particular importance are the terms of the form $G_H\varphi|H|^2$ and $g_H\varphi^2|H|^2$, where $H$ is the Higgs field, and $G_H$ and $g_H$ are the dimensionful and dimensionless couplings of the scalar $\varphi$ to the Higgs respectively.
These are the only relevant terms, i.e., terms of dimension four or lower, involving the coupling of the scalar $\varphi$ with fields of the standard model.

\begin{table}[t]
	\centering
	\begin{tabular}{ccccc}
		\hline\hline
		$\bm{T_{\rm PT}}$ & & $\bm{|G_H/M|}$ & & $\bm{|g_H|}$ \\
		\hline
		$10^7$ GeV & & $1.3\times 10^{8}$ & & $2.6\times 10^{8}$ \\
		$10^3$ GeV & & $1$ & & $1.8$ \\
		$10$ MeV & & $5.8\times 10^{-11}$ & & $8.4\times 10^{-11}$ \\
		\hline\hline
	\end{tabular}
	\caption{\small Upper bounds on the magnitude of the dimensionless coupling constants $G_H/M$ and $g_H$ assuming that they result in coupling terms that are less than a percent of the original $\varphi$ terms given in Eq.~\eqref{eHiggsTermsPot2}.  Coupling constants smaller than these will preserve the original model dynamics.}
	\label{TabHiggsCorr}
\end{table}

The $\varphi$ and $\varphi^2$ terms in the potential $U(\varphi)$ are $-\lambda M^3\varphi$ and $\lambda^2 M^2\varphi^2$, and these are modified due to the Higgs interaction as
\begin{equation}\label{eHiggsTermsPot}
    \begin{aligned}
    -\lambda M^3\varphi & \mapsto\left(-\lambda M^3+G_H|H|^2\right)\varphi,\\
    \lambda^2 M^2\varphi^2 & \mapsto\left(\lambda^2 M^2+g_H|H|^2\right)\varphi^2.
    \end{aligned}
\end{equation}

The vacuum expectation value of the Higgs field is $|H|=0$ before the electroweak phase transition and $|H|=246\,\mathrm{GeV}$ after, so these corrections apply only after this phase transition.
Since we aim to preserve the exponential form of the potential $U(\varphi)$ , we desire these additional contributions from the Higgs coupling to be small, resulting in the following conditions
\begin{equation}\label{eHiggsTermsPot2}
\left|G_H|H|^2\right|\ll|\lambda M^3|,\quad\quad\quad \left|g_H|H|^2\right|\ll\left|\lambda^2 M^2\right|,
\end{equation}
which can also be expressed as
\begin{equation}\label{eHiggsTermsPot3}
\left|\frac{G_H}{M}\right|\ll\left|\frac{\lambda M^2}{|H|^2}\right|,\quad\quad\quad \left|g_H\right|\ll\left|\frac{\lambda^2 M^2}{|H|^2}\right|.
\end{equation}

In Tab.~\ref{TabHiggsCorr}, we present the upper bounds on the magnitude of the dimensionless constants $G_H/M$ and $g_H$, assuming that they result in couplings less than a percent of the original $\varphi$ terms in Eq.~\eqref{eHiggsTermsPot2}.
We find that coupling constants smaller than those listed in the table will preserve the original model dynamics.

One can also calculate a lower bound on the couplings from the assumption that the dark sector was in thermal equilibrium and shared a common temperature with the standard model sector at some point prior to the phase transition temperature. We compute this by comparing the Hubble rate $H(T)$ at a fiducial reheating temperature of $10^{10}$~GeV with the rate of interaction between $\varphi$ and the Higgs field $H$, i.e., by requiring
\begin{equation}\label{eInteractionRates}
    \Gamma_{\varphi\leftrightarrow H}=n_\varphi\left<\sigma v\right>\geq H(T),
\end{equation}
where $\left<\sigma v\right>$ is the temperature dependent thermally averaged cross section of the interaction between $\varphi$ and $H$.
We find that for this fiducial reheating temperature that the lower bounds on the coupling constants $G_H/M$ and $g_H$, based on the interaction terms $G_H\varphi|H|^2$ and $g_H\varphi^2|H|^2$, are $8.7\times 10^{-12}$ and $2.3\times 10^{-11}$, respectively.  Comparing with Tab.~\ref{TabHiggsCorr}, we see that the upper and lower bounds on these parameters can be consistently satisfied.

\bibliographystyle{apsrev4-1}
\bibliography{ref.bib}

\end{document}